\documentclass[%
preprint,
amsmath,amssymb,
aps,showkeys,
]{revtex4-2}

\usepackage{graphicx}
\usepackage{dcolumn}
\usepackage{bm}

\graphicspath{{img/}} 
\usepackage{epsfig}
\usepackage[T2A]{fontenc}
\usepackage{latexsym}
\usepackage[utf8x]{inputenc}
\usepackage{xcolor}
\usepackage{todonotes}
\usepackage{mathtools}
\newcommand*\rfrac[2]{{}^{#1}\!/_{#2}}
\usepackage{todonotes}

\begin{document}
	
\title{Klystron type  crystal-based X-ray Volume Free Electron Laser (XVFEL)}

\author{Vladimir~G.~Baryshevsky}
\affiliation{Institute for Nuclear Problems, Minsk, Belarus}
\email{bar@inp.bsu.by}	
	
	\date{\today}

\begin{abstract}
Typical parameters of electron bunches available now  in undulator  X-ray FELs enable generation of induced radiation in crystal-based X-ray Volume free electron laser (VFEL).
An important peculiarity of an undulator  X-ray FEL is electron bunch spatial modulation with the period equal to the wavelength of the produced radiation.
Therefore, XFEL can be considered as a modulator of electron bunch density.
As a result, a facility combining an undulator, which produces a modulated beam, and a crystal, on which the modulated beam is incident, can be considered to be similar to klystron, which is a  well-known source for microwaves emission.
Use of a pre-modulated beam enables to generate and observe superradiation and induced radiation in a crystal-based X-ray VFEL.
The number of quanta, radiated in a crystal-based X-ray VFEL corresponds (or even exceeds) the number of quanta produced by XFEL in the same frequency range.	
\end{abstract}

\maketitle	
	
%
%
%
%
%
%
%
%

\section{Introduction}
\label{sec:introduction}

X-ray free electron lasers (XFELs) are the valuable tools for the scientific researchers \cite{VG_1,VG_2}.
XFELs typically employ self-amplified spontaneous emission (SASE), whereby shot noise within a high-brightness electron beam is 
amplified as it propagates through an undulator magnetic field \cite{VG_3,VG_4}.
Cavity-based X-ray FELs such as X-ray FEL oscillators (XFELO) and regenerative amplifier FEL (RAFEL) \cite{VG_5,VG_6,added_by_VG} have been proposed to produce a single temporal mode over long time.
Cavity-based FELs work by embedding an undulator within an X-ray optical cavity and returning a fraction of the radiation produced from one electron beam to seed instability for a following electron beam \cite{VG_7,VG_8,VG_9,VG_10,VG_11,VG_12,VG_13,VG_14,VG_15,VG_16,VG_17,VG_18,VG_19}.
X-ray cavities typically employ crystals acting as  Bragg  mirrors for X-ray wavelengths.
Crystal-based X-ray Volume Free Electron Laser (XVFEL) was proposed in \cite{2022_Ch_1,VG_21,VG_22,VG_23,2022_9}.

\begin{figure}[htp]
	\begin{center}	
\includegraphics[width=8cm]{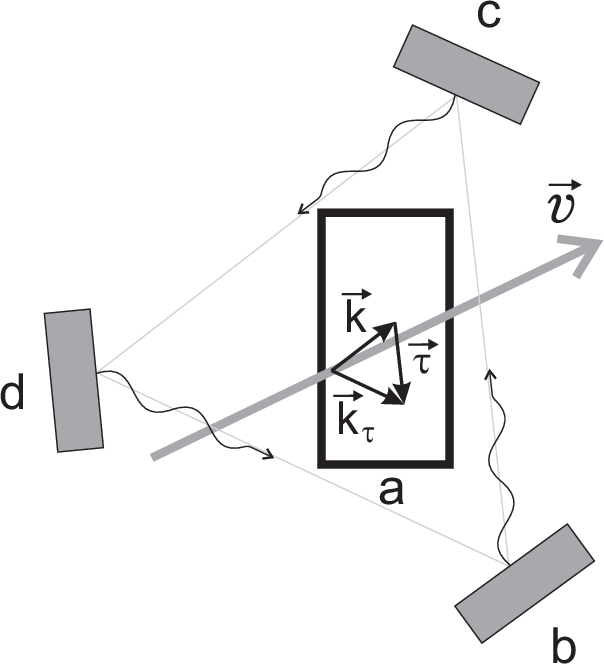}	
	\end{center}
	\caption{Crystal based X-ray Volume Free Electron Laser with Bragg reflectors comprising a crystal enclosed into a Bragg reflecting structure (optical cavity) and electron beam moving in the crystal \cite{VG_23}: a photon diffracted in crystal \textbf{a} sequentially experiences Bragg reflection from crystals \textbf{b}, \textbf{c} and \textbf{d}}
	\label{fig:fig_from_NIM}
\end{figure}

XVFEL combines the self-amplified spontaneous emission arising in a crystal due to, for example, quasi-Cherenkov parametric X-ray radiation (PXR), and the distributed feedback in a resonator established  by means of diffraction of spontaneous and induced radiation produced by the electron beam in the crystal.
%
%
Among several design options for XVFEL, there is a variant comprising a crystal enclosed into a  Bragg  reflecting structure, where an electron bunch moving in produces induced radiation \cite{VG_23} (see Fig.\ref{fig:fig_from_NIM}) i.e. according to concept 
\cite{VG_5,VG_6,VG_7,VG_8,VG_9,VG_10,VG_11,VG_12,VG_13,VG_14,VG_15,VG_16,VG_17,VG_18,VG_19}, 
XVFEL comprises an X-ray optical cavity.

Presence of  Bragg  reflectors results in significant reduction of the threshold density of the electron beam current, which is required for generation of induced radiation  \cite{VG_23}.
Therefore, in case of crystal-based X-ray Volume Free Electron Laser (XVFEL) with  Bragg  reflectors  the electron beam produces X-rays due to motion in the crystal rather then in undulator.
Using the definitions given by  
\cite{VG_5,VG_6,VG_7,VG_8,VG_9,VG_10,VG_11,VG_12,VG_13,VG_14,VG_15,VG_16,VG_17,VG_18,VG_19}
we can identify XVFEL as a cavity-based X-ray VFEL (CXVFEL).

\textcolor{black}{Present paper considers possibility to observe superradiation and induced X-ray radiation in CXVFEL, which uses an undulator X-ray FEL to provide modulation of electron bunch density.} 
%

According to analysis \cite{2022_Ch_1,VG_21,VG_22,VG_23,2022_9}
the threshold current density  of electron beam $j_{thr}$ in XVFEL is within the range $j_{thr} \approx 10^8 - 10^{10}$~A/cm$^2$, though the exact value depends on the crystal type.
The existing X-ray FELs use electron bunches comprising $N_e \sim 10^9$ electrons.
For example, according to \cite{VG_26} in LCLS XFEL for electrons with energy $E=6.7$~GeV, bunch diameter is $\sigma_b \approx 2 \cdot 10^{-5}$~cm and length is $L_b \approx 10^{-4}$~cm. 
Therefore, current density in such a bunch is as high as $j \approx 10^{14}$~A/cm$^2$. 

Consequently, typical parameters of electron bunches available now  in undulator  X-ray FELs enable generation of induced radiation in crystal-based X-ray VFEL.
An important peculiarity of an undulator  X-ray FEL is the spatial modulation of the electron bunch  with period equal to the wavelength of the produced radiation.
%
%
Therefore, XFEL can be considered as a modulator of electron bunch density.
As a result a facility combining an undulator, which produces a modulated beam, and a crystal, on which the modulated beam is incident, can be considered to be similar to klystron, which is a  well-known microwave source.
Use of pre-modulated beams eases generation and observation of induced radiation in a crystal-based X-ray VFEL \cite{2022_Ch_7,2022_Bar_NO,2022_Bar_Xray}. 

According to \cite{2022_Ch_7,2022_Bar_NO,2022_Bar_Xray}, even in case when a periodically bunched beam moving in a crystal is weakly modulated, the number of coherently emitted PXR quanta (PXR superradiance) could exceed the number of those emitted non-coherently.
	In XFEL generation of coherent X-ray radiation is accompanied by attaining high modulation for the electron bunch.
	As a result, the number of PXR superradiance quanta, produced by such a bunch moving in a crystal, appears higher.  
	According to analysis \cite{VG_28},  the number of emitted PXR superradiance quanta becomes comparable with the number of  X-ray quanta generated by XFEL, when bunch passes 1~cm length inside the crystal along the crystal surface.
	Multiple scattering of electrons by crystal atoms suppresses PXR quanta generation \cite{VG_29}, because
	it violates periodicity of beam bunching and diminishes the modulation amplitude. 
	Hence, the intensity of coherent radiation (superradiance intensity) decreases.
	It is shown hereinafter that PXR generation in  Bragg diffraction geometry enables to	
	avoid suppression of the coherent radiation by multiple scattering.
	The primary extinction, which is peculiar to Bragg diffraction geometry
	 \cite{17,18}, 
	 is responsible for exponential fading of the generated waves inside the crystal 
	 due to their diffraction even in a transparent crystal.
	The typical extinction length is $L_{ext}\approx10^{-4}$\,cm \cite{17,18}. 
	As a result, for high-energy electrons moving in a crystal (thin crystal slab) the multiple scattering does not impact the radiation generation process at the length as short as $L_{sc}\approx L_{ext}$.
\textcolor{black}{This paper is organized as follows: in Sec. II we discuss X-ray superradiation from a spatially modulated relativistic electron bunch in a crystal,
the induced X-ray radiation produced by a spatially
modulated electron bunch in crystals is considered in Sec. III, 
the concluding section is followed by Annex, which includes some useful expressions.}

\section{Coherent X-ray radiation (superradiation) from a spatially modulated relativistic bunch in a crystal}
\label{sec:1}

At the end of the long undulator in a FEL the electron beam undergoes longitudinal modulation or microbunching with the period equal to the wavelength of the FEL SASE radiation. 
When such a periodically bunched beam enters into an interaction area, superradiant emission arises (see, for instance, \cite{VG_32}). Superradiance from a single electron bunch takes place, when  bunch dimensions are smaller than the radiation wavelength (i.e. bunch duration is shorter than the period of the radiation wave). Under such conditions the intensity of radiation appears to be proportional to $N^2_{bunch}$ rather than $N_{bunch}$ ($N_{bunch}$ is the number of particles in the bunch)). Superradiance of a periodically bunched beam takes place when the beam enters into the interaction area with the rate equal to the radiation frequency. 
Superradiant emission from a bunched beam is beneficial for developing the coherent radiation sources in the wavelength range and at the operating conditions, when the accelerated beam current is too low to provide sufficient gain at a reasonable length of the interaction area. Particularly, development of the superradiant radiation sources is important for X-ray range \cite{2022_Ch_1,VG_21,2022_Ch_7,2022_Bar_NO}.

	The expression $W^s_{\vec n\omega}$ for the spectral angular distribution of radiation generated by a particle beam in a crystal reads as follows \cite{2022_Bar_NO,p11,2022_Bar_Param,VG_35}:%
	\begin{equation}\label{eq:1}
		W^s_{\vec{n}\omega}=\frac{\omega^2}{4\pi^2c^3} \left| \int\vec{E}^{s(-)*}_{\vec{k}}(\vec{r},\omega)\vec{j}(\vec{r},\omega)d^3r\right|^2,
	\end{equation}
	where $\omega$ is the frequency, $c$ is the speed of light, $s=1,2$ denotes polarization of the radiated wave, $\vec{k}=k\vec{n}$, $k=\frac{\omega}{c}$, $\vec{n}$ is a unit vector directed to the observation point,
	\begin{equation}\label{eq:2}
		\vec{j}(\vec{r},\omega)=\int e^{i\omega t}\vec{j}(\vec{r},t) dt~,
	\end{equation}
	is the Fourier transform of the beam current, which equals to
	\begin{equation}\label{eq:3}
		\vec{j}(\vec{r},t)=e\sum_l\vec{v}_{l}(t)\delta(\vec{r}-\vec{r}_{l}(t))~,
	\end{equation}
	$\vec{r}_l(t)$ is the coordinate of the $l$-th particle of the beam at the instant $t$, $\vec{v}_l(t)$ 
	is the particle velocity, $e$ is the particle electric charge, 
	$\vec{r}_l(t)=\vec{r}_{l0}+\int\limits^t_{t_0} \vec{v}_l(t')dt'$, $\vec{r}_{l0}$ is the initial coordinate of the particle i.e. particle coordinate at the instant $t=t_0$, 
	$\vec{E}_{\vec{k}}^{s(-)}$ is the solution of the homogeneous Maxwell equations including converging spherical waves; $\vec{E}^{s(-)*}_{\vec{k}}=\vec{E}^{s(+)}_{-\vec{k}}$ (see Annex~A.1).
	
	Note that the spectral-angular distribution of the number of emitted quanta  $N^s_{\vec n\omega}$ refers to $W^s_{\vec{n}\omega}$ as follows:
	\begin{equation}\label{eq:4}
		N^s_{\vec n\omega}=\frac{1}{\hbar\omega}W^s_{\vec{n}\omega}~.
	\end{equation}
	Using (\ref{eq:3}) and (\ref{eq:4}) one can express  $N^s_{\vec n\omega}$ as follows:
	\begin{equation}\label{eq:5}
		N^s_{\vec{n} \omega}=\frac{\omega e^2}{4\pi^2\hbar c^3}\left|\sum_l\int\limits_{t_0}^\infty e^{-i\omega t}\vec{v}_{l}(t)\vec{E}_{\vec k}^{s(-)}(\vec{r}_l (t), \omega) dt \right|^2.
	\end{equation}
	According to (\ref{eq:5}) the spectral-angular distribution of the emitted quanta depends on both electron trajectory $\vec{r}_l(t)$ and its velocity $\vec{v}_l(t)$.
	
	In case when no multiple scattering occurs, the coordinates and velocity for a particle steadily moving in a crystal read:
	\begin{equation}\label{eq:6}
		\vec{r}_l(t)=\vec{r}_{l0}+\vec{v}_{l0}(t-t_0), ~~\vec{v}_l(t)=\vec{v}_{l0},
	\end{equation}
	where $\vec{r}_{l0}$ and $\vec{v}_{l0}$ are the initial  coordinates and velocity of the $l$-th particle, $t_0$ is the initial instant, when  coordinate $\vec{r}_{l0}$ and velocity $\vec{v}_{l0}$ are defined (it is supposed that outside the crystal the bunch moves in vacuum), $t \ge t_0$.
	
	Electron entering into the crystal is accompanied by multiple scattering by the crystal atoms. 
	As a result, coordinate $\vec{r}_l(t)$ and velocity $\vec{v}_l(t)$ appear to be random functions as follows:
	\begin{eqnarray}\label{eq:7}
		\vec{r}_{l}(t)=\vec{r}_{l}+\vec{v}_{l0}(t-t_0)+\delta\vec{r}_{l}(t)\,, \\
		\vec{v}_{l}(t)=\vec{v}_{l0}+\Delta\vec{v}_{l}(t)\,. \nonumber
	\end{eqnarray} 
	Here $\delta\vec{r}_{l}(t)$ describes the variation of the electron trajectory caused by multiple scattering in the area occupied by the crystal; $\Delta\vec{v}_{l}(t)$ is the change of particle velocity in the crystal caused by  multiple scattering.
	
	Multiple scattering is both the reason of PXR spectral-angular distribution broadening and the origin of bremsstrahlung in the crystal.
	
	Now, let us determine the conditions under which the multiple scattering makes an impact on the intensity of coherent radiation. Note, that  $\vec{E}^{s(-)}_{\vec k}$ is expressed as a superposition of plane waves (see Annex~A.1).
	
In particular, in the area occupied by the crystal $\vec{E}^{s(-)}_{\vec k}$ can be represented as follows:
\begin{equation}\label{eq:8}
	 \vec{E}^{s(-)}_{\vec k}(\vec{r}_{l}(t),\omega)=\sum\limits_n \vec{A}_n e^{i\vec{k}_n\vec{r}_l(t)}.
\end{equation}
Let the crystal surface is parallel to (x,y) plane and $z=0$ is associated with the electron entrance into the crystal. 
Let us also define $t_0=0$. 
At this instant the electron bunch  is located in vacuum outside the crystal. 
Electron initial coordinates $\vec{r}_{l0}$ pertain to area $z<0$.
Therefore, the $l$-th electron reaches the surface $z=0$ at $t_l=|z_{l0}|/v_{zl}$. 
During this time interval the $l$-th electron shifts in the transverse direction at distance $\delta\vec{r}_{l\perp}=\vec{v}_{l\perp}|z_{l0}|/v_{zl}$. 
As a result, the transverse coordinate of the $l$-th electron, when it enters the crystal, is as follows:
\begin{equation}\label{eq:9}
	\vec{r}_{l\perp}=\vec{r}_{0\perp}+\vec{v}_{l\perp}\frac{|z_{l0}|}{v_{zl}}=\vec{r}_{l0\perp}-\vec{v}_{l0\perp}\frac{z_{l0}}{v_{zl}}\, ,
\end{equation}
while the longitudinal  coordinate of the $l$-th electron, when it enters the crystal is $z_l=0$, this occurs at the instant  $t_l=|z_{l0}|/v_{zl}$.


For high-energy particles the scattering angle $\vartheta$, considered with respect to the initial direction of particle scattering, is low: $\vartheta\ll 1, |\Delta\vec v_l|\ll\vec{v}_{l0}$. 
Therefore, the component of particle velocity $v_{l\parallel}$, which is parallel to  $\vec{v}_{l0}$, can be expressed as:
\begin{equation}\label{eq:10}
	v_{l\parallel}=v_{l0} \cos\vartheta=v_{l0}(1-\frac{\vartheta^2}{2}), ~~\Delta v_{l\parallel}=v_{l0}\frac{\vartheta^2}{2}.
\end{equation}
%
Transverse component of velocity $\vec{v}_{l\perp}$ is proportional to  $\vartheta$. 
Using the above let us consider the phase, which presents in the exponent in (\ref{eq:8}), for one of the waves:
\begin{equation}
	\varphi=\vec{k}_n\vec{r}_l(t)=\vec{k}_n \left( \vec{r}_{l0}+\vec{v}_{l0}(t-t_0)+\int\limits^t_{t_l}\Delta\vec{v}_l(t')dt'\right).
	\label{eq:11}
\end{equation}
Scattering-caused addition to phase $\varphi$ is denoted by $\delta\varphi_{sc}$ and reads as follows:
\begin{equation}
	\delta\varphi_{sc}=\vec{k}_n\int\limits^t_{t_l}\Delta\vec{v}_l(t')dt'=\vec{k}_{n\perp}\int\limits^t_{t_l}\Delta {v}_{l\perp}(t')dt'+{k}_{\parallel}\int\limits^t_{t_l}\Delta \vec{v}_{l\parallel}(t')dt'.
	\label{eq:12}
\end{equation}

If $\delta\varphi\ll1$, then multiple scattering does not contribute to phase change.

Spectral-angular distribution $N^s_{\vec n\omega}$ (see (\ref{eq:5}))
depends on the product of particle velocity  $\vec{v}_l(t)$ and polarization vector  $\vec{e}_s$ of the emitted photon, which can be expressed as follows:
%
\begin{equation}
	\vec{e}_s\vec{v}_{l}(t)=\vec{e}_s\vec{v}_{l0}+\vec{e}_s\Delta\vec{v}_l(t).
	\label{eq:13}
\end{equation}
%
%
The first term in  (\ref{eq:13}) is proportional to PXR emission angle $\vartheta_{PXR}$, while the second one depends on  scattering angle $\vartheta_{sc}$. 
Bremsstrahlung is responsible on the latter contribution 
to $N^s_{\vec{n}\omega}$. 

Detailed analysis of contributions caused by multiple scattering and bremsstrahlung to spectral-angular distribution of number of photons emitted by one electron was presented in \cite{VG_29}; comparison with experimental data was also provided therein.
Phenomenological analysis of  multiple scattering role for PXR see in \cite{2022_Bar_Param}.

Multiple scattering have no influence on spectral-angular distribution of number of quanta
 $N^s_{\vec{n}\omega}$ in case if two following inequalities are valid:
\begin{equation}\label{eq:14}
	\delta\varphi_{sc}\ll1 ~~\textrm{ and }~~ \vartheta_{sc}\ll\vartheta_{PXR}~.
\end{equation}
From (\ref{eq:12}) it follows that  $\delta\varphi_{sc}\ll1$, if two conditions are fulfilled:
\begin{equation}\label{eq:15}
	k_{n\perp}\vartheta_{sc} \ell \ll1 ~~\textrm{ and }~~ k_\parallel\vartheta_{sc}^2 \ell \ll1,
\end{equation}
where $\ell$ is the length of particle free path inside a crystal, $k_{n\perp}=|\vec{k}_{n\perp}|$, $\vec{k}_{n\perp}$ is the transversal with respect to velocity component of photon wavevector    $\vec{k}$, $k_\parallel=|\vec{k}_\parallel|$, $\vec{k}_\parallel$ is the parallel  with respect to velocity component of photon wavevector  $\vec{k}$.

For evaluations $\vartheta_{sc}$ is supposed to be the typical angle of scattering, which is determined by the mean-square angle of particle multiple scattering in crystal $<\vartheta^2>$, i.e. \begin{eqnarray*}
	\vartheta_{sc}=\sqrt{<\vartheta^2>}\,.
\end{eqnarray*}
The mean-square angle of particle multiple scattering can be expressed as:
\begin{equation}\label{eq:16}
	<\vartheta^2>=g\frac{E^2_s}{E^2}\frac{l}{L_{rad}}\, ,
\end{equation}
where $E_s=21$~MeV, $E$ is the particle energy, $L_{rad}$ is the radiation length, $g$ is the coefficient, which describes the difference between the mean square angle of multiple scattering $<\vartheta^2>$  for crystal   and that for  amorphous medium (in case of amorphous medium $g=1$).
 When an electron moves at a small angle relatively to the crystal axes, the magnitude of $g$ can be much greater than unity.
Therefore, we can rewrite (\ref{eq:15}) in the following way:
\begin{equation}\label{eq:17}
	k_{n\perp}\sqrt{<\vartheta^2>} \ell \ll1 \textrm{ and } k_\parallel<\vartheta^2> \ell \ll1\,.
\end{equation}
The first part in (\ref{eq:17}) results in:
\begin{equation}\label{eq:18}
	\ell \ll\left( \frac{\lambda}{2\pi}\frac{mc^2}{E_s}\right)^{\rfrac{2}{3}}L_{ray}^{\rfrac{1}{3}}\gamma^{\frac{4}{3}} \,,
\end{equation}
where $\lambda$ is the radiation wavelength; $k_{n\perp}\sim\frac{k}{\gamma}$, here $\frac{1}{\gamma}$ is the typical angle of scattering for a relativistic particle.
Second part of (\ref{eq:17}) gives:
\begin{equation}\label{eq:19}
	\ell \ll \left( \frac{\lambda}{2\pi}L_{rad}\right)^{\rfrac{1}{2}}\left( \frac{mc^2}{E_s}\right) \gamma \,.
\end{equation}
From (\ref{eq:18}), (\ref{eq:19}) we can, for example,  obtain that  in $Si$ crystal for electrons with Lorentz-factor $\gamma=10^4$ both evaluations give $\ell$ to be smaller than $10^{-1} - 10^{-2}$~cm, when X-ray photon wavelength is as short as $\lambda=10^{-8} - 10^{-9}$~cm.


For a thick crystal multiple scattering significantly contributes to the spectral-angular distributions of emitted photons \cite{VG_29}.
For further consideration let us suppose that 
crystal thickness coincides with $\ell$ and the above conditions 
ensure weak effect of 
 multiple scattering on PXR emission.

Note, that electrons of the bunch entering the crystal have some velocity spread $\delta v$. 
Bunches produced in X-ray FEL typically have 
$\frac{\delta v_\perp}{v}\ll\frac{1}{\gamma}$ and $\left( \frac{\delta v_\parallel}{v}\right) \ll\frac{1}{\gamma^2}$
($\delta v_\perp$ is the spread in transverse velocity (transverse with respect to the average velocity), $\delta v_\parallel=v-v_\parallel$ is the spread in velocity parallel  to the average velocity.

Analyzing (\ref{eq:5}), (\ref{eq:8}) one can get the conditions, under which the beam velocity spread $\delta v_l$ can be neglected: to achieve this the longitudinal dimensions $L_b$ of the bunch should satisfy the relations 
\begin{equation}\label{eq:20}
	\frac{\omega}{v}\frac{\delta v_\perp}{v}L_b < 1 \textrm{ ~~and~~ } k_{\perp}\frac{\delta v_{\perp}}{v}L_b < 1.
\end{equation}
i.e.
\begin{equation}\label{eq:21}
	L_b <\frac{1}{k(\delta\vartheta)^2} \textrm{ ~~and~~ } L_b <\frac{1}{k\vartheta_\gamma\delta\vartheta},
\end{equation}
where $\delta\vartheta$ is the characteristic angular of particles velocities in the bunch; $\vartheta_\gamma\le\frac{1}{\gamma}$ is the angle of photon emission.

According to  \cite{VG_26,VG_28}, the typical angular spread of velocities in the bunch is $\delta\vartheta\sim 10^{-4}~$rad, bunch length is $L_b=10^{-4}$~cm. 
Therefore, for X-ray wavelengths, when  $\lambda\approx10^{-8}-10^{-9}$~cm, requirements (\ref{eq:21}) are fulfilled. 
Crystal is supposed to be thin, to satisfy conditions (\ref{eq:18}), (\ref{eq:19}) for the bunch.

As a result, to obtain the expression for spectral-angular distribution of radiation produced by the bunch, it is sufficient to average (\ref{eq:5}) over the distribution of electron coordinates in the bunch.


Thus, from equation (\ref{eq:5}) we have the following expression for spectral-angular distribution of the number of quanta emitted  by the bunch \cite{2022_Bar_NO,2022_Bar_Xray}:
\begin{equation}\label{eq:22}
	\frac{d^2N}{d\omega d\Omega}=\frac{d^2N_1}{d\omega d\Omega}N_e+\frac{d^2N_1}{d\omega d\Omega}\sum\limits_{l=m}\left\langle e^{-i\vec{K}\vec{r}_l}e^{i\vec{K}\vec{r}_m}\right\rangle,
\end{equation}
where $N_e$ is the number of particles in the bunch,
\begin{equation*}
	\vec{r}_l=(\vec{r}_{0l\perp}-\vec{v}_\perp\frac{z_{0l}}{v_z},\frac{v}{v_z}z_{0l}); \vec{K}=(\vec{k}_\perp,\frac{\omega}{v}).
\end{equation*}
Brackets <...> mean averaging over distribution of electron coordinates in the bunch.
Since spread in electron velocities can be neglected,  then
the above mentioned averaging  turns  to averaging over the initial distribution of coordinates $\vec{r}_{0l}$ of the particle in the bunch.

Let us denote the function of electrons distribution over coordinates in the bunch as $w(\vec{r})$; 
 $\int w(\vec{r})d^3r=1$.
The explicit expression for this function, which is used to describe generaion of X-rays in X-ray FEL (see, for example, \cite{VG_26,VG_28}) can be wtitten as follows:
\begin{equation}\label{eq:23}
	w(\vec{r})=\frac{1}{\pi\sigma^2_b}e^{-\frac{x^2+y^2}{\sigma^2_b}}\frac{1}{N_b}\sum\limits^{N_b}_{p=0}\frac{1}{\sqrt{\pi}\sigma_c}e^{\rfrac{-(z-pd_0)^2}{\sigma_c^2}},
\end{equation}
where $\sigma_b$ is the typical transverse size of the bunch, $d_0$ is the period of microbunches location, 
$L_b=N_b d_0$ is the bunch length, fluctuations of bunch period are described by parameter  $\sigma_c\ll d_0$, $N_b$ is the number of microbunches in the bunch.
Using (\ref{eq:23}) spectral-angular distribution (\ref{eq:22}) can be re-written as follows:
\begin{eqnarray}\label{eq:24}
	\frac{d^2N}{d\omega d\Omega}=\frac{d^2N_1}{d\omega d\Omega}\left( 1-\left| \int e^{-i\vec{K}\vec{r}}w(\vec{r})d^3r\right|^2\right) N_e \nonumber \\
	+\frac{d^2N_1}{d\omega d\Omega}\left| \int e^{-i\vec{K}\vec{r}}w(\vec{r})d^3r\right|^2 N_e^2\,.
\end{eqnarray}
The first term is proportional to the number of electrons in the bunch and describes incoherent radiation from the bunch. 
The second term is proportional to $N_e^2$ and describes coherent spontaneous radiation (superradiation) produced by the bunch.

Bunch form-factor (integral  in (\ref{eq:23})) is the Fourier-transform for $w(\vec{r})$ and can be explicitly expressed as follows (see also \cite{VG_28}):
\begin{equation}\label{eq:25}
	F(\vec{K})=\int e^{-i\vec{K}\vec{r}}w(\vec{r})d^3r
	= e^{-k^2_\perp\frac{\sigma_b^2}{4}} \cdot \frac{d_0(1-e^{iL_b\frac{\omega}{v}})}{L_b(1-e^{id_0\frac{\omega}{v}})}\cdot e^{-\frac{\omega^2}{v^2}\frac{\sigma_c^2}{4}}\,.
\end{equation}

According to  \cite{VG_26,VG_28} for LCLS XFEL facility $\gamma \approx 10^4,$ parameter $\sigma_b=2\cdot10^{-5}$~cm,  number of electrons $N_e\approx10^9$, bunch length $L_b=10^{-4}$~cm, microbunches period $d_0=10^{-8}$~cm, $\sigma_c=10^{-9}$~cm i.e. $\sigma_c\ll d_0$.

Further analysis requires consideration of explicit expressions for  $\frac{d^2N_1}{d\omega d\Omega}$: see, for example, in   \cite{2022_Bar_NO,p11,2022_Bar_Param}
those describing PXR radiated at both large and small angles with respect to the  particle velocity direction.
%
%
\textcolor{black}{
	PXR should be analyzed for several geometry layouts, namely: Laue geometry, Bragg geometry and surface diffraction case \cite{VG_21,2022_Bar_NO,p11,2022_Bar_Param,VG_35,VG_36}.}
%
\textcolor{black}{
	Analysis provided hereinafter demonstrates that in Bragg case influence of multiple scattering on superradiation and induced PXR emission could be sufficiently suppressed.} 
	
\begin{figure}[h]
	\epsfxsize = 6 cm
	\centerline{\epsfbox{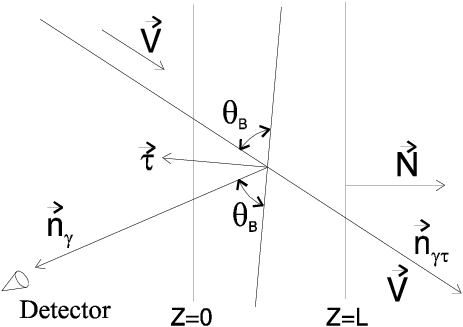}}
	\caption{Scheme of diffraction geometry: 
		diffraction maximum is expected to be observed at large angle equal to $2\theta _{B}$ (here $\theta _{B}$ is  Bragg 
		angle, $\sin \theta _{B}=\frac{|\vec{v}\vec{\tau}|}{\tau }$) with respect to direction of
		particle velocity  in  Bragg  geometry}
	\label{fig:2}
\end{figure}

\begin{figure}[ht]
	\centering
	\epsfxsize = 6 cm
	\centerline{\epsfbox{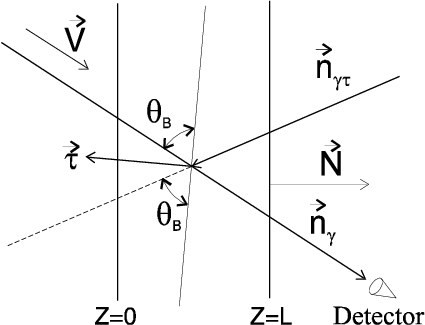}}
	\caption{Scheme of Bragg  diffraction in the forward direction: 
		diffraction maximum is expected to be observed at a small angle with respect to
		particle velocity direction, Bragg condition is fulfilled}
	\label{fig:FBD}
\end{figure}

\textcolor{black}{Let us consider the Bragg case. One can observe an electromagnetic
wave emitted by a charged particle in the direction of diffraction, which is leaving
the crystal through that surface, through which the particle enters into
the crystal (see Fig.\ref{fig:2}). 
Bragg case is also actualized, when radiation leaves
crystal through the surface, through which the particle exits from the
crystal slab i.e. surface located at $z = L$ (see Fig.\ref{fig:FBD}).}

For Bragg  diffraction in the forward direction expression for  $\frac{d^2 N^s_f}{d \omega d\Omega }$  reads as follows:
%
\begin{equation}\label{eq:26}
	\frac{d^2 N^s_f}{d \omega d\Omega }=\frac {e^2 \omega }{4\pi^2
		\hbar c^3}(\vec{e}_s \vec{v})^2 \times
	\left| \sum_{\mu =1,2}\gamma _{\mu s}^{0} e^{i\frac{\omega }{c\gamma _{0}}
		\varepsilon _{\mu s}L}\left[ \frac{1}{\omega -\vec{k}\vec{v}}-\frac{1}
	{\omega -\vec{k}_{\mu s}\vec{v}}\right]
	\left[ e^{i(\omega -\vec{k}_{\mu s}\vec{v})\frac{L}
		{c\gamma _{0}}}-1\right] \right| ^{2}
\end{equation}
$$\gamma_{1(2)s}^0 =
[2\varepsilon_{2(1)s}-\chi_0]
\left[(2\varepsilon_{2(1)s}-\chi_0)-(2\varepsilon_{1(2)s}-\chi_0)\exp\left[i\frac{\omega}{\gamma_0}(\varepsilon_{2(1)s}-
\varepsilon_{1(2)s})L\right]\right]^{-1}.$$
For  Bragg  case in the diffraction direction 
\begin{equation}\label{eq:27}
	\frac{d^{2}N^s_d}{d\omega d \Omega }=\frac{e^{2}\omega
	}{4\pi ^{2}\hbar c^{3}}(\vec{e}_{\tau s}\vec{v})^{2}\times \left|
	\sum_{\mu =1,2}\gamma _{\mu s}^{\tau }\left[ \frac{1}{\omega
		-\vec{k}_{\tau } \vec{v}}-\frac{1}{\omega -\vec{k}_{\mu \tau
			s}\vec{v}}\right] \left[ e^{i(\omega -\vec{k}_{\mu \tau
			s}\vec{v})\frac{L}{c\gamma _{0}}}-1\right] \right| ^{2},
\end{equation}
$\vec{k}_{\mu s}=\vec{k}+\frac{\omega }{c}\varepsilon _{\mu s}^*\vec{N}$, $\mu =1,2$, 
$\vec{N}$ is the unit vector perpendicular to the crystal surface and directed inside it,
%
%
%

$$\gamma _{1(2)s}^{\tau } =-\beta _{1}[C_{s}\chi _{\tau }]
\left[(2\varepsilon _{2(1)s}-\chi _{0})-(2\varepsilon _{1(2)s}-\chi _{0})\exp \left[i
\frac{\omega }{c\gamma _{0}}(\varepsilon _{2(1)s}-\varepsilon _{1(2)s})L\right]\right]^{-1},$$

\begin{equation}\label{eq:28}
	\varepsilon_{1(2)s}=
	\frac{1}{4}\left[(1-\beta_{1})\chi_{0}^s + \alpha _{B}\beta _{1}\right] 
	\pm \frac{1}{4}\left\{ \left[(1+\beta_{1})\chi_{0}^s -\alpha _{B}\beta _{1}\right]^2
	+4\beta _{1}C^2_s\chi_{\tau }^s\chi_{-\tau }^s\right\}^{\rfrac{1}{2}},
\end{equation}
$\alpha _{B}=(2\vec{k}\vec{\tau}+\tau^2)k^{-2}$,  $\gamma _{0}=\vec{n}_{\gamma }\vec{N}$, $\vec{n}_{\gamma }=\frac{\vec{k}}{k}$, $\beta _{1}=\frac{\gamma_{0}}{\gamma _{1}}$,$\gamma _{1}=\vec{
	n}_{\gamma \tau }\cdot\vec{N}$, $\vec{n%
}_{\gamma \tau }=\frac{\vec{k}+\vec{\tau}}{|\vec{k}+\vec{\tau}|}$,
%
%
  $C_{s}=\vec{e}_{s}\vec{e}_{\tau s}$ , $\vec{e}_{s}$ $(\vec{e}_{\tau s})$ are
the unit polarization vectors of the incident and diffracted waves,
respectively.

Let us consider expressions (\ref{eq:26}) and (\ref{eq:27}) in more details. 
Polarizabilities $\chi(\vec{\tau})$ in (\ref{eq:26}) and (\ref{eq:27}) have both real and imaginary parts. 
Imaginary part of $\chi(\vec{\tau})$ is associated with X-ray quanta absorption in matter.
If X-ray frequency is higher as compared to the typical atomic frequencies, then $\textrm{ Re } \chi(\tau) \gg \textrm{ Im }\chi(\tau)$ \cite{17}.

For example, for  $Si$ crystal and X-ray quanta with the energy in the range $10 - 30$~keV,  the real part of polarizability  is as high as  $\textrm{ Re } \chi(\tau)\approx10^{-5} - 10^{-6}$, while the 
imaginary part of $\chi(\tau)$ is much lower: $\textrm{ Im } \chi(\tau)\approx3\cdot10^{-7} - 5\cdot10^{-9}$. 
Therefore, for X-ray quanta  of energy $10$~keV  absorption length  $L_{abs}\approx6\cdot10^{-3}$~cm and for that  of energy $30$~keV $L_{abs}\approx10^{-1}$~cm.

Quanta absorption in a crystal limits growth of PXR intensity with the increase of crystal thickness. 
That is why the geometry implying electron motion along the crystal surface (either in vacuum or inside crystal) 
was proposed in  \cite{VG_21,VG_28,VG_36}: 
diffracted radiation is observed  in vacuum at some angle to crystal surface. 
Thus, the whole length of electron path in the crystal contributes to the intensity of radiation observed in vacuum. 
This length could be much longer than the length of X-ray quanta absorption in the crystal.

However, hereinafter it was mentioned that the longer is the length of electron path in the crystal, the stronger is dependence of radiation  spectral-angular distribution on multiple scattering of particles in the crystal.
Theory of multiple scattering influence on PXR and comparison of theoretical analysis with experimental data were given in  \cite{VG_29}. 
It was demonstrated that multiple scattering significantly contributes to the spectral-angular distributions of PXR emitted photons and should be considered, when interpreting experimental data.

Mind that for Bragg  and Laue diffraction cases multiple scattering influence on PXR could be different.
Let us analyze value of  $\varepsilon_{1(2)s}$ in  expressions (\ref{eq:26}), (\ref{eq:27}) for  Bragg  and Laue geometries.

For Laue geometry  $\beta_{1}=\frac{\gamma_{0}}{\gamma_{1}}>0$. 
Therefore, the radicand in expression for $\varepsilon_{1(2)s}$ (\ref{eq:28}) includes sum of two terms.
In case, if frequency of X-ray radiation exceeds typical atomic frequencies,
$\textrm{ Re }\chi_{\vec{\tau}}^s\gg \textrm{ Im }\chi_{\vec{\tau}}^s$.
Let us neglect the imaginary part of 
 $\chi_{\vec{\tau}}$\,; in this case, expressions for $\varepsilon_{1(2)s}$ become purely real.
Wave vectors comprising in exponents in expressions for $\frac{d^2 N}{d \omega d\Omega }$
and $\vec{E}(\vec{r},\omega)$ become purely real, too, i.e. waves are not damping in the 
crystal.

For  Bragg  geometry parameter $\beta_{1}=\frac{\gamma_{0}}{\gamma_{1}}<0$\,.
Therefore, radicand in  $\varepsilon_{1(2)s}$ comprises difference of two terms. 
If neglect $\textrm{ Im } \chi(\tau)$, then radicand
includes difference of two positive real terms as follows:
\begin{equation}\label{eq:29}
	\varepsilon_{1(2)s}=
	\frac{1}{4}\left[(1+|\beta_{1}|)\chi_{0}^s -\alpha_{B}|\beta_{1}|\right] 
	\pm \frac{1}{4}\left\{ \left[(1+|\beta_{1}|)\chi_{0}^s -\alpha_{B}|\beta_{1}|\right]^2
	-4|\beta _{1}|C^2_s\chi_{\tau }^s\chi_{-\tau }^s\right\}^{\rfrac{1}{2}}.
\end{equation}
Therefore, if the following inequality is valid
\begin{equation}\label{eq:30}
	\left[ (1+|\beta_{1}|)\chi_{0}^s -\alpha_{B}|\beta_{1}|\right]^2 < 4|\beta _{1}|C^2_s\chi_{\tau }^s\chi_{-\tau }^s,
\end{equation}
then radicand is negative and $\varepsilon_{1(2)s}$ is complex:
\begin{equation}\label{eq:31}
	\varepsilon_{1(2)s}=
	\frac{1}{4}\left[(1+|\beta_{1}|)\chi_{0}^s -\alpha_{B}|\beta_{1}|\right] 
	\pm \frac{i}{4}\left\{4|\beta_{1}|C^2_s\chi_{\tau }^s\chi_{-\tau }^s-\left[(1+|\beta_{1}|)\chi_{0}^s -\alpha_{B}|\beta_{1}|\right]^2\right\}^{\rfrac{1}{2}}.	
\end{equation}

Wave vectors comprising in exponents in $E^s_k$ and $\frac{d^2 N}{d \omega d\Omega }$ become complex. 
Waves both damping and growing appear inside the  
crystal.

	
 Using boundary conditions one can obtain the expression for $\vec{E}$, which is valid both in vacuum and inside the crystal. 
 Therefore, if condition (\ref{eq:30}) holds, then the wave incident on the crystal appears totally reflected, while the wave traveling inwards the crystal is exponentially damping. 
 %
%
Thus, the primary extinction arises  \cite{17,18}. 
%
According to (\ref{eq:31})  the imaginary part of $\varepsilon_{1(2)s}$ achieves its maximal value when the following equality is valid
\begin{equation}\label{eq:32}
	(1+|\beta_{1}|)\chi_0^s-\alpha_{B}|\beta_{1}|=0 \,.
\end{equation}
Mind that parameter $\alpha_{B}=\frac{2\vec{k}\vec{\tau}+\tau^2}{k^2}$.
Therefore, $\alpha_{B}=\frac{2k\tau \cos \vartheta_{k\tau}+\tau^2}{k^2}$, $\vartheta_{k\tau}$ is the angle between wave vector
$\vec{k}$ and reciprocal lattice vector $\vec{\tau}$.
Angle $\vartheta_{k\tau}$ can be expressed as  $\vartheta_{k\tau}=\vartheta_{k\tau}^{Br}+\Delta\vartheta_{m}$,  where 
$\vartheta_{k\tau}^{Br}$ is the angle between $\vec{k}$ and $\vec{\tau}$ in case of  Bragg diffraction. If one neglect the influence of refraction on wave propagation in a crystal, then $\vartheta_{k\tau}^{Br}=\frac{\pi}{2}+\vartheta_{Br}$, $\vartheta_{Br}$ is the angle  
 of Bragg diffraction, i.e. angle complying with Bragg condition: $2\sin\vartheta_{Br}=\frac{\tau}{k}$, i.e. $2\sin\vartheta_{Br}=\frac{n\lambda}{d}, 
n=1,2..., \lambda$ is the wavelength, $d$ is the distance  between reflecting crystallographic planes.

In X-ray range, for which quanta frequencies exceed typical atomic frequencies  
$\chi_0^s(\chi_{\tau }^s)<0$ and 
$|\chi_0^s|(|\chi_{\tau }^s|)\ll1$. 
That is why    correction  $\Delta\vartheta_{m}$, which is caused by wave refraction, is small $\Delta\vartheta_{m}\ll1$, when 
the imaginary part of $\varepsilon_{1(2)s}$ achieves its maximal value.
In this case parameter $\alpha_{B}$ can be written as follows:
\begin{equation}\label{eq:33}
	\alpha_{B}=-\frac{2\tau}{k}\Delta\vartheta_{m}\cos\vartheta_{Br}+\frac{\tau^2}{2k^2}(\Delta\vartheta_{m})^2.
\end{equation}
Note, that in expression  (\ref{eq:33}) we keep the second-order term over $(\Delta\vartheta_{m})^2$.
This is made because in the range of angles  
 $\vartheta_{Br}$ close to $\frac{\pi}{2}$, the term proportional to  $\Delta\vartheta_{m}$ tends to zero.
Using (\ref{eq:33}) equation (\ref{eq:32}) can be expressed as follows:
\begin{equation}\label{eq:34}
	(\Delta\vartheta_{m})^2-2\cot\vartheta_{Br}\cdot\Delta\vartheta_{m}-\frac{1}{2\sin^2\vartheta_{Br}}\frac{1+|\beta_{1}|}{|\beta_{1}|}\chi_0^s=0.
\end{equation}
%
%
Equation  (\ref{eq:34}) has the following roots:
\begin{equation}\label{eq:35}
	\Delta\vartheta_{m1(2)}=\cot\vartheta_{Br} \left(1\pm\sqrt{1-\frac{1}{2\cos^2\vartheta_{Br}}\frac{(1+|\beta_{1}|)}{|\beta_{1}|}|\chi_0^s|}\right).
\end{equation}
Mind that in X-ray range 
$\textrm{Re} \chi_0^s<0, \textrm{ Im } \chi_0^s\ll |\textrm{Re}\chi_0^s|, \chi_0^s\approx-|\chi_0^s|$.
Since $|\chi_0^s|$ is small at angles  $\vartheta_{Br}\ll\frac{\pi}{2}$, 
the second term in radicand (\ref{eq:35}) is much smaller then 1 in this case.
Therefore, small-parameter expansion over this term can be carried. 
As a result,  one can get the following expression for deviation from angle 
$\vartheta_{k\tau}$ due to wave refraction  $\Delta\vartheta_{m2}$, corresponding to sign (-) in (\ref{eq:35}):
\begin{equation}\label{eq:36}
	\Delta\vartheta_{m2}=\frac{1}{2\sin2\vartheta_{Br}}\frac{(1+|\beta_{1}|)}{|\beta_{1}|}|\chi_0^s|,
\end{equation}
i.e, (see, for example, (\ref{eq:30}) and (\ref{eq:31}))
\begin{equation}\label{eq:37}
	\Delta\vartheta_{m2}=\frac{1}{2\sin2\vartheta_{Br}}\left(1+\frac{|\gamma_{1}|}{|\gamma_{0}|}\right)|\chi_0^s|.
\end{equation}
Root $\Delta\vartheta_{m1}$, corresponding to sign (+) in (\ref{eq:35}) should be omitted, because at 
$\vartheta_{Br}\ll\frac{\pi}{2}$ it appears to be much greater than 1 and becomes nonphysical.

In case of symmetric reflection, when the absorption length caused by primary extinction is minimal, the angle between $\vec{k}$ and $\vec{\tau}$ reads as $\vartheta_{k\tau}=\vartheta_{Br}+\Delta\vartheta_{m2}$ and differs from  $\vartheta_{Br}$ by addition as follows:
\begin{equation}\label{eq:38}
	\Delta\vartheta_{m2}=\frac{1}{sin2\vartheta_{Br}}|\chi_0^s|.
\end{equation}
If $\vartheta_{Br}$ tends to $\frac{\pi}{2}$, then roots of (\ref{eq:35}) converge towards each other and at $\cos^2\vartheta_{Br}=\frac{1}{2}\frac{(1+|\beta_{1})}{|\beta_{1}}|\chi_0^s|$ they coincide i.e. $\Delta\vartheta_{m1}=\Delta\vartheta_{m2}=\Delta\vartheta_{m}$. Deviation $\Delta\vartheta_{m}$ in this case reads as follows:
\begin{equation}\label{eq:33_2}
	\Delta\vartheta_{m}=\frac{1}{\sqrt{2}}\sqrt{\frac{(1+|\beta_{1}|)}{|\beta_{1}|}}\sqrt{|\chi_0^s|}.
\end{equation}
In  symmetric case $\Delta \vartheta_m=\sqrt{|\chi_0^s|}$.

Let us define now the boundaries of total reflection (boundaries of primary extinction); 
they are determined by angle deviations
 $\Delta\vartheta_{m1(2)}$, enabling square root in (\ref{eq:29}) and (\ref{eq:31}) to vanish.

Since $|\chi_0^s|$ and $|\chi_{\tau }^s|$ are low at angles $\vartheta_{Br}\ll\frac{\pi}{2}$, then the angular width $\Delta\Theta$ of the total reflection area is determined by the difference in deviations $\Delta\vartheta_{m1}$ and $\Delta\vartheta_{m2}$, and can be expressed as follows:
\begin{equation}\label{eq:34_2}
	\Delta\Theta=\frac{2|C_s|}{\sin2\vartheta_{Br}}\sqrt{\frac{\gamma_{1}}{\gamma_{0}}}|\chi_{\tau }^s|.
\end{equation}
Formula (\ref{eq:34_2}) was obtained very long ago by Laue, Ewald and Darwin, who all contributed to the dynamical diffraction theory (see, for example, \cite{17,18}).

According to (\ref{eq:34_2}) with $\vartheta_{Br}$ growth and tending $\vartheta_{Br}$ to $\frac{\pi}{2}$,  the value of  $\Delta\Theta$ also grows and tends to infinity.
Analysis demonstrates that in this case deviations  $\vartheta_{m1(2)}$ are described by the following expression:
\begin{equation}\label{eq:35_2}
	\Delta\vartheta_{m1(2)}=\sqrt{\frac{1}{2}\frac{(1+|\beta_{1}|)}{|\beta_{1}|}|\chi_0^s|\pm\frac{1}{\sqrt{|\beta_{1}|}}|\chi_{\tau }^s|};
\end{equation}
for symmetric reflection $|\beta_{1}|=1.$

Therefore, the width of reflection area, i.e.  the area, where primary extinction exists, can be expressed as:
\begin{equation}\label{eq:36_2}
	\Delta\Theta=\sqrt{|\chi_0^s|+|\chi_{\tau }^s|}-\sqrt{|\chi_0^s|-|\chi_{\tau }^s|}.
\end{equation}
In case if
\begin{equation}\label{eq:37_2}
	|\chi_{\tau }^s|\approx|\chi_0^s|\textrm{, then } \Delta\Theta=\sqrt{2|\chi_{\tau }|}.
\end{equation}
In case if $|\chi_{\tau }|\ll|\chi_0|$, then

\begin{equation}\label{eq:38_2}
	\Delta\Theta=\frac{|\chi_{\tau }|}{\sqrt{|\chi_0|}}\gg|\chi_{\tau }|\textrm{, because }\sqrt{\chi_0}\ll 1.
\end{equation}
Therefore, for Bragg reflection at $\vartheta_{Br}\approx\frac{\pi}{2}$ the angular width in the area of total reflection is significantly wider than in the area, where $\vartheta_{Br}\ll\frac{\pi}{2}$.

It should be reminded that in the range of  total reflection the phenomenon of primary extinction arises that results in exponential decay of the wave, which is the sum (interference) of initial and diffracted waves, inside the crystal. 
Absorption length  $\l_{ext}$ is determined by the  imaginary part of  $\varepsilon_{1(2)s}$ (см. (\ref{eq:31})) and, for example, in case of symmetric Bragg reflection in the range of angles  $\vartheta_{Br}\sim \frac{\pi}{2}$ it can be expressed as follows:
\begin{equation}\label{eq:39}
	l_{ext}=\frac{1}{k\,\,\textrm{Im}\varepsilon_{1(2)s}}\sim\frac{1}{k|\chi_{\tau }|}\sim\frac{\lambda}{2\pi|\chi_{\tau }|},
\end{equation}
where $\lambda$ is the wavelength.

For wavelength range $\lambda\approx10^{-8} - 10^{-9}$~cm and $\chi_{\tau }\approx10^{-5} - 10^{-6}$, length $l_{ext}$ is small: $l_{ext}\approx10^{-4} - 10^{-3}$~cm.
At this length multiple scattering does not influence on PXR (see   evaluations   (\ref{eq:18}), (\ref{eq:19}) hereinabove).

Therefore, multiple scattering of electrons in a crystal in Bragg geometry in the range of angles $\vartheta_{Br}$ close to $\frac{\pi}{2}$ could be ignored (in this case mind to provide electron beam motion out of the angles close to Lindhard angle to avoid influence of channeling on electron scattering).

From the expression  for $\varepsilon_{1,2,s}$ it also follows that that frequency range $\Delta\omega$, for which total Bragg reflection is possible, is described by inequality $\Delta\omega\approx|\chi_{\tau }|\omega$.

Let us consider PXR in Bragg geometry \cite{p11}.
If particle velocity is directed at the angle close to  $\frac{\pi}{2}$ to the crystal surface, then PXR photons leave the crystal along the velocity direction and form the extinction wave damping inside the crystal moving deeper in it (in the direction opposite to particle velocity).
Extinction length in this case is
 $l_{ext}^{out}\sim\frac{\lambda}{2\pi|\chi_{\tau }|}$, i.e., 
$l_{ext}^{out}\sim10^{-4} - 10^{-3}$~cm.

Therefore, when an electron passes a crystal slab of thickness  $L$, two radiation pulses arise: one  directed along the velocity, when the particle leaves the crystal, and another one directed oppositely,  when the particle enters the crystal.
%
%
%
%
%
%
This is similar to convenient transition radiation formed by a particle passing through a slab of matter.
However, in this case radiation formation is significantly different. 

Total reflection, which is typical for PXR in Bragg geometry, is caused by Bragg diffraction and interference of two waves.
Transition radiation arises due to leap of refraction index at the boundary vacuum--matter (matter--vacuum)
(for more details see Annex~A.2). 
Moreover, 
%
for X-ray range 
 in case if a thin crystal slab has thickness  $L\sim L_{ext}$, 
 conventional transition radiation is suppressed, because 
thickness $L$ is much shorter than coherent radiation length  $l_{coh}\sim\lambda\gamma^2$. 
For $\lambda\sim10^{-8} - 10^{-9}$~cm and $\gamma\sim 10^4 $ coherent radiation length $l_{coh}\sim10^{-1} - 1$~cm and $\delta n\ll 1$.
Therefore, use of thin slabs with $L\sim L_{ext}$ to generate PXR superradiation  enables to study this effect in case,  when both multiple scattering and transition radiation are suppressed.

Let us use (\ref{eq:22})-(\ref{eq:25}) to analyse the above. 
From expression (\ref{eq:27}) for $\frac{d^2 N_1}{d \omega d\Omega }$ the following expression for number of quanta produced by a single electron in symmetric Bragg case in the range of angles $\vartheta_{Br}\sim\frac{\pi}{2}$ at crystal  slab  thickness $L\ge L_{ext}$:
\begin{equation}\label{eq:40}
	\frac{d^2 N_1}{d \omega d\Omega }\approx \frac{e^2\omega}{4\pi^2\hbar c^3}(\vec{e}_{s \tau} \vec{v})^2 
	\left| \frac{1}{\omega-\vec{k}_{\tau}\vec{v}}-\frac{1}{\omega-\vec{k}_{2\tau s}\vec{v}}\right| ^2\,.
\end{equation}
From (\ref{eq:24}), (\ref{eq:25}) it follows that the number of PXR superaradiance quanta is expressed as follows:

\begin{equation}\label{eq:41}
	N_{super}\approx N_e^2\iint\frac{d^2 N_1}{d \omega d\Omega } \left|F(\vec{K})\right|^2 d \omega d\Omega\,.
\end{equation}
According to (\ref{eq:25}) form-factor $F(\vec{K})$ reads:

\begin{equation*}
	F(\vec{K})=e^{\rfrac{-k_\perp^2\sigma_b^2}{4}}\frac{d_0}{L_b}\frac{(1-e^{iL_b\frac{\omega}{v}})}{(1-e^{id_0\frac{\omega}{v}})}e^{\rfrac{-\frac{\omega^2}{v^2}\sigma_c^2}{4}}\,.
\end{equation*}
Therefore,
\begin{eqnarray*}
	|F(\vec{K})|^2 &=& e^{\rfrac{-k_\perp^2\sigma_b^2}{2}}\left|\frac{d_0}{L_b}\frac{(1-e^{iL_b\frac{\omega}{v}})}
	{1-ie^{id_0\frac{\omega}{v}}}\right|^2 e^{\rfrac{-\frac{\omega^2}{v^2}\sigma_c^2}{4}} \\
	&\approx&
	e^{\rfrac{-k_\perp^2\sigma_b^2}{2}}\cdot\frac{2\pi v}{L_b}\sum\limits_{n=1}^{\infty}\delta(\omega-\frac{2\pi v}{d_0}n)e^{\rfrac{-\frac{\omega^2}{v^2}\sigma_c^2}{4}}\,.
\end{eqnarray*}
%
%
%
Making evaluations keep in mind that bandwidth $\Delta\omega_{ext}$, within which the total reflection takes place, is $\frac{\Delta\omega_{ext}}{\omega}\approx\chi_{\tau }^s$.
Let us compare this width with the bandwidth of function $F(\vec{K})$.
Length of the periodically modulated bunch is $L_b$.
 Therefore, bunch passes though the crystal  boundary during $\Delta t_b=\frac{L_b}{v}$. 
 This time determines the initial duration of time interval during which radiation is excited inside the crystal. 
 Therefore, the uncertainty in superradiation frequency  $\Delta\omega=\frac{2\pi}{\Delta t_0}=\frac{2\pi v}{L_b}, L_b=d_0 N_b$, $N$ is the number of periods. 
The same result follows from the expression for bunch form-factor:
\begin{equation}\label{eq:42}
	|F(\vec{K})|^2=e^{\rfrac{-k_\perp^2\sigma_b^2}{2}}\frac{d_0^2}{L_b^2}\frac{\sin^2L_b\frac{\omega}{2v}}{\sin^2d_0\frac{\omega}{2v}}e^{\rfrac{-\frac{\omega^2}{v^2}\sigma_c^2}{4}} \,.
\end{equation}
From $|F(\vec{K})|^2$ one can obtain that for frequencies $\omega_r=\frac{2\pi v}{d_0}n, n=0, 1, 2...$ the following ratios are valid $\frac{\sin^2L_b\frac{\omega}{2v}}{\sin^2d_0\frac{\omega}{2v}}=N_b^2$. 
When $\omega$ differs from $\omega_r$ by $\Delta\omega=\frac{2\pi v}{L_n}$ the expression $\sin L_b\frac{\omega}{2v}$ vanishes for the first time. 
Therefore, values
 $\Delta\omega=\frac{2\pi v}{L_n}$ 
 can be considered as a typical bandwidth of $F(\vec{K})$ distribution over the frequency.

Let us compare bandwidth $\Delta\omega=\frac{2\pi v}{d_0}$ with that for reflection, for which primary extinction presents i.e. for $\Delta\omega_{ext}\approx\chi_{\tau }^s\omega$.
%
%
According to the example in  \cite{VG_28}, for XFEL radiation frequency is $\omega\approx2\cdot10^{19}$~s$^{-1}$. Therefore, $\Delta\omega_{ext}\approx10^{-5}\cdot2\cdot10^{19}\approx2\cdot10^{14}$~s$^{-1}$, 
at bunch length $L_b\approx10^{-4}$~cm value
$ \Delta\omega_r\approx\frac{2\pi v}{L_b}\approx2\cdot10^{15}$, i.e. $\Delta\omega_r\gg\Delta\omega_{ext}$.

Let us consider distribution over angles $\vartheta$, which is determined by form-factor $F(\vec{K})$.
According to (\ref{eq:42}) it is given by parameter $k_\perp\sigma_b=k\vartheta\sigma_b$.
When $k_\perp\sigma_b>1$, the distribution goes down quickly.
Therefore,  $\vartheta\sim\frac{1}{k\sigma_b}$ can be considered as a typical angle.
According to \cite{VG_26,VG_28}, $\sigma_b=2\cdot10^{-5}$~cm,
therefore, $\vartheta\approx\frac{c}{\omega\sigma_b}\approx10^{-4}$~rad.
Angular width $\frac{d^2N_1}{d\omega d\Omega}$, within which primary extinction exists, is determined by  angular width
$\Delta\vartheta_{ext}\approx\sqrt{\chi_{\tau }^s}$, i.e. value
$\Delta\vartheta_{ext}\approx 10^{-2} - 10^{-3}$. 
Therefore, one can integrate (\ref{eq:41}) 
over $d\Omega$ in the range, where  the quantity  $\frac{d^2N_1}{d\omega d\Omega}$ is constant.

Integration of exponent $e^{\rfrac{-k_\perp^2\sigma_b^2}{2}}$
%
%
 over $d\Omega$ leads to the following expression
\begin{equation*}
	J=\int e^{\rfrac{-k_\perp^2\sigma_b^2}{2}} d\Omega = \frac{4\pi}{k^2\sigma_b^2} \,.
\end{equation*}

Therefore, for number of superradiant quanta  $N_{coh}$ the following evaluation can be obtained:
\begin{equation}\label{eq:43}
	N_{coh}=\frac{d^2N_1}{d\omega d\Omega}|_{\omega=\frac{2\pi vn}{d_0}}\cdot\frac{4\pi}{k^2\sigma_b^2}|\chi_{\tau }|\,\omega \, e^{-\frac{\pi^2 n^2}{d_0^2}\sigma_c^2}N_e^2.
\end{equation}
Frequency $\omega$ is in the vicinity of Bragg reflection condition with accuracy $\Delta\omega\sim|\chi_{\tau }|\omega$.

Due to presence of form-factor $F(\vec{K})$, typical  angle $\vartheta$ between the direction of the wavevector of coherently radiated photon  and the particle velocity can be  evaluated as follows: $\vartheta\sim\frac{1}{k\sigma_b}\approx10^{-4}$~rad.
Frequency range  $\Delta\omega$ in the vicinity of Bragg reflection freqency $\omega_{B}$, within which primary extinction exists, reads $\Delta\omega\sim|\chi_{\tau }|\omega_{B}$.
According to the above consideration and using (\ref{eq:40}) one can evaluate spectral-angular distribution 
 $\frac{d^2N_1}{d\omega d\Omega}$ in the range, where primary extinction presents:
\begin{equation}\label{eq:44}
	\frac{d^2N_1}{d\omega d\Omega}|_{\omega=\omega_{B}}\approx\frac{\alpha}{\pi^2\omega}\vartheta^2\left|\frac{1}{\gamma^2+\vartheta^2}\right|^2, 
\end{equation}
where $\alpha=\frac{e^2}{\hbar c}$ is the fine structure constant.

Therefore, for the number of superradiation photons  the following evaluation is valid:
\begin{equation}\label{eq:44_2}
	N_{coh}\approx\frac{4\alpha}{\pi}\vartheta^4\left| \frac{1}{\gamma^2+\vartheta^2}\right| ^2 |\chi_{\tau }|e^{-\frac{\pi^2 n^2}{d_0^2}\sigma_c^2}N_e^2 \,.
\end{equation}
If Lorentz-factor is  $\gamma \ge 10^4, \vartheta\sim10^{-4}$ and $n=1, \sigma_c\ll d_0$, then $N_{coh}\approx\frac{4\alpha}{\pi}
|\chi_{\tau }|N_e^2$.

From (\ref{eq:44_2}) the number of superradiation quanta in Bragg reflection geometry  is $N_{coh}\approx10^{11}$, if number of electrons in bunch  $N_e\approx10^9$. 
Quanta are radiated within typical angle $\vartheta\sim10^{-4}$
and spread in frequencies $\frac{\Delta\omega}{\omega}\approx10^{-5}$ that corresponds (or even exceeds) the number of quanta produced by XFEL in the same frequencey range  $\frac{\Delta\omega}{\omega}\approx10^{-5}$ (according to \cite{VG_26} for XFEL coherent radiation  $\frac{\Delta\omega}{\omega}\approx10^{-3}$).

%
%
%

\section{Induced X-ray radiation produced by a spatially modulated electron bunch in a crystal. Klystron type of XVFEL}

An electron bunch, which produces PXR superradiation pulse in a crystal, is affected by X-ray radiation produced by the electron beam in XFLEL undulator.  
The PXR superradiation pulse also affects the electron bunch. 
Therefore, induced radiation is produced along with spontaneous radiation.
Radiation absorption by the electron beam also influences the number of the induced radiation quanta.


It is well-known that probabilities of emission of spontaneous and stimulated (induced) radiation are described by Einstein coefficients that can be used to describe amplification of intensity of radiation emitted by beam particles 
%
%
\cite{2022_Ginzburg_theor, 2022_Ginzburg_Tsyt, 2022_Fedorov, 2022_Marshall}. 

Applying quantum conservation laws: laws of momentum and energy conservation, one can calculate corrections for quantum recoil. 
These corrections cause shift of spectral-angular distributions associated with photon emission by a particle with respect to distributions associated with photon absorption by the particle.
%
Therefore, probabilities of spontaneous emission and absorption are also different for the emitted and absorbed photons, which frequencies are equal: this corresponds to the equal angle between photon emission direction and initial momentum of the particle and that between momentum of absorbed photon  and initial momentum of the particle.
%

Quantum conditions for stimulated amplification of radiation from the particle beam have been studied for a beam moving in a boundless medium. 
%
%
Analysis presented hereinafter is given for emission and absorption of radiation produced  by a bunch for both cases: a boundless and bounded media.
%

%
%

Since correction for quantum recoil is small, then emission and absorption probability can be expanded over this correction as a small parameter. Thus, analysis of radiation amplification can be reduced to calculating the derivative of probability of spontaneous emission \cite{2022_Ginzburg_theor, 2022_Ginzburg_Tsyt, 2022_Fedorov, 2022_Marshall}. 

\subsection{Quantum recoil for Cherenkov radiation in uniform medium}


Analysis of induced radiation from a bunch requires consideration of conservation laws. Let us consider a bunch  moving in a uniform medim. 
Energy and momentum conservation laws in case of emission (absorption) of a photon reads:
%
%
\begin{equation}
	\label{eq:66}
	E_0=E_1 \pm \hbar \omega ; ~~
	\vec{p}_0=\vec{p}_1 \pm \hbar \vec{k},
\end{equation}
where $E_0$, $\vec{p}_0$ are the energy and momentum of a particle before emission (absorption) of a photon, $\hbar \vec{k}$ is the momentum of the photon; 
$ \hbar \vec{k}= \frac{\hbar \omega}{c} n_l(\omega, \vec{s}) \vec{s}$, photon frequency is denoted by $\omega$, 
$n_l(\omega, \vec{s})$ is the refraction index for the wave of type $l$, which propagates in the considered medium, $\vec{s}$ is the unit vector along the propagation direction, $c$ is the speed of light.
Mind that particle energy and moment relate as  $ E_{0,1} = \sqrt{p_{0,1}^2c^2 + m^2c^4} $, where $m$ is the particle mass.
%

Using (\ref{eq:66}) one can get the expression \cite{2022_Ginzburg_theor} for angle $\vartheta_\gamma$ between direction of 
emission the photon with frequency $\omega$  and initial momentum of the particle: 
\begin{equation}
	\label{eq:67}
	\cos \vartheta_{\gamma}^{rad}= \frac{c}{vn_l}\left[ 1 + \frac{\hbar \omega_{rad}}{2E} (n_l^2-1) \right],
\end{equation}
\begin{equation}
	\label{eq:68}
	\hbar \omega_{rad}= \frac{2E}{cn_l (1-\frac{1}{n_l^2})} (v \cos\vartheta_{\gamma}^{rad} - \frac{c}{n_l}).
\end{equation}
Using the similar approach one can obtain the expression for angle  $\vartheta_\gamma^{absorb}$ (the angle between  photon momentum and initial momentum of the particle) for the absorbed photon with frequency $\omega$ as follows:

\begin{equation}
	\label{eq:69}
	\cos \vartheta_{\gamma}^{absorb}= \frac{c}{vn_l}\left[ 1 - \frac{\hbar \omega_{absorb}}{2E} (n_l^2-1) \right],
\end{equation}

\begin{equation}
	\label{eq:70}
	\hbar \omega_{absorb}= \frac{2E}{cn_l (1-\frac{1}{n_l^2})} (v \cos\vartheta_{\gamma}^{absorb} - \frac{c}{n_l}).
\end{equation}
Note that for classic consideration with the use of  Maxwell equations $\vartheta_{\gamma}^{rad}$ is defined by:
\begin{equation}
	\label{eq:71}
	\cos\vartheta_{\gamma}^{rad}= \frac{c}{vn_l}.
\end{equation}
Therefore, consideration of  quantum recoil for photon emission described by  (\ref{eq:67})-(\ref{eq:68}) results in small correction to expression (\ref{eq:71}). 
%
%
For case of photon absorption according to (\ref{eq:69}), (\ref{eq:70}) the similar addition caused by quantum recoil also appears. 
From the above consideration it follows that angles
$\vartheta_{\gamma}^{rad}$ and $\vartheta_{\gamma}^{absorb}$ differ from each other, as well as frequencies $\omega_{rad}$ and $\omega_{absorb}$. 
Therefore, probability of radiation and probability of absorption reach maximal value at different conditions
\cite{2022_Ginzburg_theor, 2022_Ginzburg_Tsyt, 2022_Fedorov, 2022_Marshall}.

\subsection{Quantum recoil for Cherenkov (quasi-Cherenkov) radiation in a parallel-sided slab}

Presence of boundaries causes photon wavefunction conversion from the plane wave to the superposition of waves, which momentum changes due to refraction by the boundary.
%
%
Therefore conservation laws read as follows:

\begin{equation}
	\label{eq:72}
	E_0=E_1 \pm \hbar \omega, ~~
	p_{0z}=p_{1z} \pm \hbar {k_z},~~ \vec{p}_{0\perp}=\vec{p}_{1\perp} \pm \hbar \vec{k}_{0 \perp},
\end{equation}
where $k_z=\sqrt{k_0^2 n_l^2 - k_{0 \perp}^2}$, $ k_0=\frac{\omega}{c}$, axis $z$ is orthogonal to the slab surface, $(x,y)$ plane coincides with the slab surface, $k_{0 \perp}^2=k_{0x}^2+k_{0y}^2$,
$k_{0 \perp}^2=k_{0}^2 \sin^2 \vartheta$, $\vartheta$ is the angle between $\vec{k}_0$ and unit vector $\vec{s}_z$ (directed along $z$) in vacuum. 
Vector component $k_z$ of the wavevector $\vec{k}$ reads as follows:
\begin{equation}
	\label{} 
	k_z = \frac{\omega}{c} \sqrt{n_l^2 - \frac{k_{0 \perp}^2}{k_{0}^2}}=\frac{\omega}{c}\sqrt{n_l^2 - \sin^2 \vartheta}.
	\nonumber
\end{equation}
Let us analyze the following expression:
\begin{equation}
	\label{eq:73}
	p_{0z} - p_{1z} \mp \hbar {k_z}=0
\end{equation}
using $p_{0z}= p_0 \cos \vartheta_0, p_{1z}= p_1 \cos \vartheta_1$, $\vartheta_0 (\vartheta_1)$ is the angle between  
$\vec{p}_{0z} (\vec{p}_{1z})$ and $\vec{s}_z$.
Let us assume that $ \vartheta_0, \vartheta_1 \ll 1 $, i.e. the incident particle enters the plane approximately normally. 
Recall that
\begin{eqnarray}
	\label{}
	p_0 &=& \frac{1}{c} \sqrt{E^2 -m^2c^4},\\
	p_1 &=& \frac{1}{c} \sqrt{E_1^2 -m^2c^4}=\frac{1}{c} \sqrt{(E \mp \hbar \omega)^2 -m^2c^4}. \nonumber
\end{eqnarray}
Suppose that particle energy $E$ is greater than photon energy $\hbar \omega$  and use $\hbar \omega \ll E$ to obtain the following expression:
\begin{eqnarray}
	\label{eq:74*}
	p_1 &=& \frac{1}{c} \sqrt{E^2 \mp 2E \hbar \omega +\hbar^2 \omega^2 - m^2 c^4} \\
	&=& p_0 \sqrt{1 \mp \frac{2E \hbar \omega}{p_{0}^{2}c^2} +\frac{\hbar^2 \omega^2}{p_{0}^{2}c^2}} , \nonumber
\end{eqnarray}
i.e.
\begin{equation}
	\label{eq:74}
	p_1 \simeq p_0 \left( 1 \mp \frac{2E \hbar \omega}{p_{0}^{2}c^2} + \frac{\hbar^2 \omega^2}{2p_{0}^{2}c^2} - \frac{1}{8} \frac{4 E^2 \hbar^2 \omega^2}{p_{0}^{4}c^4}\right).
\end{equation}
Component $p_{1z}$ reads:
\begin{eqnarray}
	\label{eq:75}
	p_{1z} &=& p_1 \sqrt{1 - \frac{p^2_{1 \perp}}{p^2_{\perp}}} \simeq p_1 \left( 1 - \frac{p^2_{1 \perp}}{2 p^2_{1}} \right) \nonumber \\
	&=& 
	p_1 \left(1 - \frac{p^2_{0 \perp} \mp 2 \vec{p}_{0 \perp}\hbar \vec{k}_{0 \perp} +\hbar^2 k^2_{\perp}}{2 p_1^2} \right), \nonumber \\ 
	p_1^2 &=& p_0^2 \mp \frac{2E \hbar \omega}{c^2}+\frac{\hbar^2 \omega^2}{c^2}\,.
\end{eqnarray}
If the initial momentum of the particle is directed along  $ z $ (normal incidence), then $  \vec{p}_{0 \perp} = 0 $. 
Therefore:
\begin{eqnarray}
	\label{eq:76}
	p_{1z} &=& p_1 - \frac{\hbar^2 k^2_{\perp}}{2p_1} \simeq \nonumber \\ &\simeq& p_0 \left(1 \mp \frac{E \hbar \omega}{p_{0}^{2}c^2} + \frac{\hbar^2 \omega^2}{2p_{0}^{2}c^2} -\frac{1}{2} \frac{E^2 \hbar^2 \omega^2}{p_{0}^{4}c^4} \right) - \frac{\hbar^2 k^2_{ 0 \perp}}{2p_0}= \nonumber \\ 
	&=& p_0 \mp \frac{\hbar \omega}{v} + \frac{\hbar^2 \omega^2}{2p_{0}c^2} - \frac{1}{2} \frac{\hbar^2 \omega^2}{p_0 v^2} - \frac{\hbar^2 k^2_{\perp}}{2 p_0}= \\ \nonumber
	&=& p_0 \mp \frac{\hbar \omega}{v} - \frac{\hbar^2 \omega^2}{2p_0 v^2} \left(1 - \frac{v^2}{c^2} \cos^2 \vartheta \right).
\end{eqnarray}
Therefore (\ref{eq:73}) can be expressed as follows:
\begin{eqnarray}
	\label{eq:77*}
	\pm \frac{\hbar \omega}{v} + \frac{\hbar^2 \omega^2}{2p_{0}v^2} \left( 1 - \frac{v^2}{c^2} \cos^2 \vartheta \right) \mp \frac{\hbar \omega}{c} \sqrt{n_l^2 - \sin^2 \vartheta}=0, \nonumber
\end{eqnarray}
i.e.
\begin{eqnarray}
	\label{eq:77}
	\pm \left( 1- \frac{v}{c} \sqrt{n_l^2 - \sin^2 \vartheta}\right) + \frac{\hbar \omega}{2p_0v} \left( 1- \frac{v^2}{c^2} \cos^2 \vartheta  \right) = 0.
\end{eqnarray}
Vavilov-Cherenkov condition for a slab reads:
%
$ 1- \frac{v}{c} \sqrt{n_l^2- \sin^2 \vartheta}=0 $.
Therefore, 
$\frac{v^2}{c^2} (n_l^2 - \sin^2 \vartheta) =1$, 
i.e. $\cos^2 \vartheta + (n_l^2-1) =\frac{c^2}{v^2}$.
Thus,  $ 1-\frac{v^2}{c^2}\cos^2 \vartheta = \frac{v^2}{c^2} (n_l^2-1) $
and for emission and absorption (\ref{eq:77}) gives the following: 
%
\begin{eqnarray}
	\label{eq:78}
	\text{for emission~~~~~\,} 1-\frac{v}{c} \sqrt{n_l^2 - \sin^2 \vartheta} + \frac{v~\hbar \omega (n_l^2-1)}{2p_0c^2}=0 , \\ \nonumber
	\text{for absorption~~~} 1-\frac{v}{c} \sqrt{n_l^2 - \sin^2 \vartheta} - \frac{v~\hbar \omega  (n_l^2-1)}{2p_0c^2}=0.
\end{eqnarray}
Calculating frequency $\omega$ from (\ref{eq:78}) one can get the frequency shift due to quantum recoil effects.
%
%
For relativistic case $ \frac{v}{c} $ in the third summand of (\ref{eq:78}) can be set $ \frac{v}{c}=1$. 
%

When studying quantum recoil effect for quasi-Cherenkov radiation one should replace  momentum $\hbar 
\vec{k}$ in (\ref{eq:72}) by the photon momentum in a spatially periodic medium i.e. by photon wave vector in a spatially periodic medium $ \hbar \vec{k}_l $; the expressions for  $ \hbar \vec{k}_l $ are obtained in 
%
\cite{2022_Bar_NO,p11, 2022_Bar_Param}.

\subsection{Quantum recoil for PXR in  Bragg  case}

Let us now consider correction for quantum recoil for PXR in the range of parameters, for which primary extinction presents.
%
In contrast to Vavilov-Cherenkov effect in this case the conservation laws read as follows: 
%

\begin{equation}
	\label{eq:79}
	E_0=E_1 \pm \hbar \omega, ~~ \vec{p}_0 =\vec{p}_1 \pm \hbar \vec{k} + \vec{q},
\end{equation}
where $\vec{q}$ ~is the transmitted momentum, $ \vec{k} = \frac{\omega}{c} \vec{s}_{\gamma} $.
Note that expressions (\ref{eq:79}) are also valid for  bremsstrahlung.
%
%
According to (\ref{eq:79}) the transmitted momentum can be expressed as follows: 
\begin{equation}
	\label{eq:80}
	\vec{q}=\vec{p}_0-\vec{p}_1 \mp \hbar \vec{k}\,.
\end{equation}
Multiplying $\vec{q}$ ~by unit vector $\vec{s}$ directed along the initial particle momentum $\vec{p}_0={p}_0\vec{s}$ one can get the following expression for transmitted momentum component parallel to the initial particle momentum:
\begin{equation}
	\label{eq:81}
	{q}_{\parallel}={p}_0-{p}_1 \cos \vartheta_1 \mp \hbar \frac{\omega}{c} \cos \vartheta_{\gamma},
\end{equation}
where $\vartheta_{1}$ is the angle between momenta $ \vec{p}_0 $ and $ \vec{p}_1 $; $ \vartheta_{\gamma} $ is the angle between momenta  $\vec{p}_0$ and $\vec{k}$.

Typically, momentum $\hbar \vec{k}$ of a photon produced by transition radiation  is  smaller than initial particle momentum  $\vec{p}_0$. Therefore $\vartheta_{1} \ll 1$ and the following expression is valid:
\begin{equation}
	\label{eq:82}
	{q}_{\parallel}={p}_0 - {p}_1 - {p}_1 \frac{\vartheta_1^2}{2} \mp \frac{\hbar\omega}{c} \cos \vartheta_{\gamma}\,.
\end{equation}
Momentum ${p}_1$ is expressed as follows (see (\ref{eq:74})):
\begin{eqnarray}
	\label{eq:82*}
	p_1 &=& \frac{1}{c}\sqrt{(E \mp \hbar \omega)^2 -m^2c^4} =  \\ 
	&=& p_0\left( 1 \mp \frac{E \hbar \omega}{p_0^2c^2} + \frac{\hbar^2 \omega^2}{2p_0^2c^2} - \frac{1}{2} \frac{E^2 \hbar^2 \omega^2}{p_0^4c^4}\right). \nonumber
\end{eqnarray}
Projection $p_1 \cos \vartheta_1 $ is expressed by (\ref{eq:76}), $z$ is directed along $ \vec{p}_0 $. Therefore:
\begin{equation}	
	\label{eq:83}
	{q}_{\parallel}= \pm \frac{\hbar \omega}{v} + \frac{\hbar^2 \omega^2}{2p_0v^2} \left( 1- \frac{v^2}{c^2} \cos^2 \vartheta_{\gamma}\right) \mp \hbar \frac{\omega}{c} \cos \vartheta_{\gamma}\,,
\end{equation}
i.e.
\begin{equation}	
	\label{eq:84}
	{q}_{\parallel}= \frac{\hbar \omega}{v}\left(\pm (1- \frac{v}{c} \cos \vartheta_{\gamma}) +  \frac{\hbar \omega}{2p_0v} (1-\frac{v^2}{c^2} \cos^2 \vartheta_{\gamma})\right)\,,
\end{equation}
\begin{equation}	
	\label{eq:85}
	|q|_{\parallel}= \frac{\hbar \omega}{v}\left( (1- \frac{v}{c} \cos \vartheta_{\gamma}) \pm \frac{\hbar \omega}{2p_0v} (1-\frac{v^2}{c^2} \cos^2 \vartheta_{\gamma}) \right).
\end{equation}
According to (\ref{eq:84}), (\ref{eq:85}) the expression for longitudinal incident momentum includes quantum correction with a sign different for emitted and absorbed photons. 
As a result, the probabilities of emission and absorption reach their maximal values at different $q_{\parallel}$.  


Since correction is small, then both emission and absorption probabilities can be expanded over this correction as a small parameter. 
Thus, analysis of radiation amplification can be reduced to calculation of derivatives $ \Delta \vartheta_{\gamma} $ and $ \Delta \omega $ (i.e. $d\vartheta_{\gamma}$ and $d \omega$)  of probability of spontaneous radiation (superradiation),
%
%
%
and then one should average the obtained expressions with the distribution of particle momentum
\cite{2022_Ginzburg_theor, 2022_Ginzburg_Tsyt, 2022_Fedorov, 2022_Marshall}. 



Therefore, it is fundamentally important that analyzing probabilities of photon emission and absorption by an electron beam one should consider quantum recoil, which an electron experiences when either emitting or absorbing a photon.
%
%
When emitting a photon the electron is slowing down, while when a photon is absorbed the electron is accelerated.
%
%
If quantum recoil is not considered, then the line shapes for emission and absorption coinside; 
radiation is not absorbed by the bunch.
%
%

Quantum recoil phenomenon leads to shift of frequencies associated with emission and absorption lines with respect to their classic values.
%
%
Exactly this effect enables amplification of radiation by the electron beam   \cite{2022_Ginzburg_theor, 2022_Ginzburg_Tsyt, 2022_Fedorov, 2022_Marshall}. 
The difference in emission and absorption lines caused by a moderate frequency shift makes the amplification coefficient proportional to the derivative of the spontaneous emission line shape \cite{2022_Ginzburg_theor, 2022_Ginzburg_Tsyt, 2022_Fedorov, 2022_Marshall}.
%
%
The same is valid for induced PXR radiation in crystal-based XVFEL  \cite{VG_22,VG_23}.

According to analysis \cite{2022_Ch_1,VG_22,VG_23,2022_Ch_7} the required current density for the electron bunch  to achieve generation threshold in XVFEL is $j\approx10^8 - 10^{10}~\frac{A}{cm^2}$.
%
%
According to the above analysis, current densities for electron bunches available in XFEL reach values $j\approx10^{13} - 10^{14}~\frac{A}{cm^2}$. 
%
%
This provides possibility to observe generation of  superradiation and induced quasi-Cherenkov PXR in crystal-based XVFEL.

Let us emphasize to conclude that klystron type of XVFEL %
	{see Fig. \ref{fig:fig_from_NIM}}
 can be realized   in  XFEL with an optical cavity \cite{VG_5,VG_6,VG_8}: if the electron bunch passes through the crystal of the optical cavity rather than diverting outside the  crystal (in full accordance with the approach used in  \cite{VG_5,VG_6,VG_8}).
Number of superradiation quanta in Bragg reflection geometry can be evaluated using (\ref{eq:44_2}) to be as high as $N_{coh} \approx 10^{11}$ for number of electrons in the bunch $N_e \approx 10^9$. 
Quanta are radiated within typical angle $\vartheta \sim 10^{-4}$ with spread in frequencies $\Delta \omega / \omega \approx 10^{-5}$ that corresponds (or even exceeds) the number of quanta produced by XFEL in the same frequency range $\Delta \omega / \omega \approx 10^{-5}$ (compare with $\Delta \omega / \omega$ for XFEL coherent radiation, which is approximately equal to  $ 10^{-3}$ ).

\section*{Conclusion}

Typical parameters of electron bunches available now  in undulator  X-ray FELs enable generation of induced radiation in crystal-based X-ray VFEL.
An important peculiarity of an undulator  X-ray FEL is spatial modulation of the electron bunch  with the period equal to the wavelength of the produced radiation.
Therefore, XFEL can be considered as a modulator of electron bunch density.
As a result, a facility combining an undulator, which produces a modulated beam, and a crystal, on which the modulated beam is incident, can be considered to be similar to klystron, which is a  well-known source for microwaves emission.
Use of pre-modulated beams enables to generate and observe superradiation and induced radiation in a crystal-based X-ray VFEL.

\newpage

\bigskip

\section*{Annex A}
\label{annex:A}

Refraction of electromagnetic field associated with an electron passing through a matter with a uniform velocity originates both Vavilov-Cherenkov and transition radiations. Cherenkov emission of
a photon by a charged particle occurs whenever the index of refraction $%
n(\omega )>1$, $\omega $ is the photon frequency. 
For X-ray frequencies exceeding atomic typical frequencies the refractive index is generally presented in the following form:
\begin{equation}
	n(\omega )=1-\frac{\omega _{L}^{2}}{2\omega ^{2}}\,,
\end{equation}
where $\omega _{L}$ is the Langmuir frequency.

As it was first shown in \cite{1} 
 the crystal
structure essentially changes the refraction index for virtual X-ray photons
emitted under diffraction conditions. 
This leads to change of the generation conditions for bremsstrahlung and Cherenkov radiation. 
As a result, when a
charged particle is moving with a constant velocity it emits
quasi-Cherenkov X-ray radiation \cite{1,2}.
 Moreover, photons emited by the particle can be observed 
 not only at a small angle to the direction of particle motion ($\vartheta \sim \gamma ^{-2}$, here
$\gamma $ is Lorentz factor of the particle), but also at
much larger angles and
even in the backward direction. 
Namely this fact allowed to observe parametric X-ray radiation (PXR) for the first time
\cite{3,4,4a}:
the detected radiation peaks were shown  to
be actually a new type of radiation rather than diffracted bremsstrahlung
radiation \cite{3,4,4a,5}.

In accordance with  
\cite{2022_Ch_7,2022_Bar_NO,2022_Bar_Param,p11,1}
 diffraction of virtual photons in a crystal 
can be described by the set of
refraction indices $n_{\mu }(\omega,\vec{k})$, some of which
may appear to be greater than unit. 
Here $n(\omega ,\vec{k})$ is the refraction index of the crystal for X-ray photons
with wave vector $\vec{k}$. 
Particularly, in case of two-wave
diffraction the refraction index is greater then unit $n_1 (\omega ,\vec{k})>1$ and
$n_2 (\omega ,\vec{k})<1$. 
Therefore, two waves propagate in the crystal: the slow
wave with $n_{1}>1$ and the fast wave with $n_{2}<1$, respectively.

\subsection*{A.1. General expressions for PXR emission rate}

Let a particle moving with a uniform velocity be incident on a crystal slab
with thickness  $L<<L_{br}$,   where
\ $L_{br}=(\frac{\omega q}{c})^{-1/2}$\ is the coherent length for bremsstrahlung radiation, $q=\overline{\theta }%
_{s}^{2}/4$\ and $\overline{\theta }_{s}^{2}$ is a root-mean-square angle of
multiple scattering for charged particles per unit length. 
This
requirement allows us to neglect the multiple scattering of particles by
atoms of crystal. 
The method for theoretical analysis in case of intense 
multiple scattering is described
in \cite{VG_29}.

The expression for the differential number of photons emitted in direction $\vec{k}$ with polarization
vector $\vec{e}_{s}$ is as follows (see, for example, \cite{2022_Bar_NO,2022_Bar_Param,p11}):
\begin{equation}
	\frac{d^{2}N_{s}}{d\omega d\Omega }=\frac{e^{2}Q^{2}\omega }{4\pi ^{2}\hbar
		c^{3}}\left| \int\limits_{-\infty }^{+\infty }\vec{v}\vec{E}_{\vec{k}%
	}^{(-)s}(\vec{r}(t),\omega )\exp (-i\omega t)dt\right| ^{2},
\label{eq:99_3}
\end{equation}
where $eQ$ is the particle charge, $\vec{E}_{\vec{k}}^{(-)s}$\ is the
solution of homogeneous Maxwell equation. In order to determine the number
of quanta emitted by a particle passing through a crystal slab one should
find the explicit expressions for the solutions $\vec{E}_{\vec{k}}^{(-)s}$. 
The field $\vec{E}_{\vec{k}}^{(-)s}$ can be derived from
relation $\vec{E}_{-\vec{k}}^{(-)s}=(\vec{E}_{\vec{k}}^{(+)s})^{\ast }$\
using solution $\vec{E}_{\vec{k}}^{(+)s}$, which describes the ordinary
problem of photon scattering by a crystal.

The following set of equations  for   wave amplitudes
in case of two-beam diffraction can be obtained:
\begin{eqnarray}
	\left( \frac{k^{2}}{\omega ^{2}}-1-\chi _{0}^{\ast }\right) \vec{E}_{\vec{k}%
	}^{(-)s}-C_{s}\chi _{-\tau }^{\ast }\vec{E}_{\vec{k}_{\tau }}^{(-)s} &=&0 \,,
	\nonumber \\
	\left( \frac{k^{2}}{\omega ^{2}}-1-\chi _{0}^{\ast }\right) \vec{E}_{\vec{k}%
	}^{(-)s}-C_{s}\chi _{-\tau }^{\ast }\vec{E}_{\vec{k}_{\tau }}^{(-)s} &=&0 \,.
\label{eq:99_4}
\end{eqnarray}
Here $\vec{k}_{\tau }=\vec{k}+\vec{\tau}$ , $\vec{\tau}$ is the reciprocal
lattice vector, $|\tau |=2\pi /d,$ $\ d$ is the interplanar distance, $\chi
_{0},\chi _{\tau }^{s},\chi _{-\tau }^{s}$ are the Fourier components of the
crystal susceptibility, $C_{s}=\vec{e}_{s}\vec{e}_{\tau s}$ , $\vec{e}_{s}$ $(\vec{e}_{\tau s})$ are
the unit polarization vectors of the incident and diffracted waves,
respectively. It is well known that a crystal is described by a
periodic susceptibility (see, for example \cite{17,18}):
\begin{equation}
	\chi (\vec{r})=\sum_{\tau }\chi _{\tau }\exp (i\vec{\tau}\vec{r}).
\label{eq:99_5}
\end{equation}
The condition for the linear system (\ref{eq:99_4}) to be solvable leads to the dispersion
equation, which determines  wave vectors $\vec{k}$ existing in a crystal.
It is convenient to present these wave vectors in the form$\ \ \ \vec{k}%
_{\mu s}=\vec{k}+\kappa _{\mu s}\vec{N},$  where $\kappa _{\mu s}=\frac{%
	\omega }{c\gamma _{0}}\varepsilon _{\mu s},$  $\mu =1,2$; $\vec{N}$ is
the unit vector normal to the entrance surface of the crystal slab and directed inward the crystal.

The general solution of equation (\ref{eq:99_4}) inside the crystal is as follows:
\begin{equation}
	\vec{E}_{\vec{k}}^{(-)s}(\vec{r})=\sum_{\mu =1}^{2}\left[ \vec{e}^{s}A_{\mu
	}\exp (i\vec{k}_{\mu s}\vec{r})+\vec{e}_{\tau }^{s}A_{\tau \mu }\exp (i\vec
{k}_{\mu s\tau }\vec{r})\right].
	\label{eq:99_8}
\end{equation}
Matching these solutions with the solutions of Maxwell equation for the
vacuum area we can find the explicit form of $\vec{E}_{\vec{k}}^{(-)s}(\vec{r%
})$ throughout the space; different 
diffraction geometries should be considered, namely, Laue and  Bragg  diffraction.

The differential number of PXR quanta
with polarization vector
$\vec{e}_s$ diffracted in forward direction for the  Bragg  case reads as follows:
\begin{equation}
	\frac{d^2 N_s}{d \omega d\Omega }=\frac {e^2 Q^2 \omega }{4\pi^2
		\hbar c^3}(\vec{e}_s \vec{v})^2
	\left| \sum_{\mu =1,2}\gamma _{\mu s}^{0} e^{i\frac{\omega }{c\gamma _{0}}
		\varepsilon _{\mu s}L}\left[ \frac{1}{\omega -\vec{k}\vec{v}}-\frac{1}
	{\omega -\vec{k}_{\mu s}\vec{v}}\right]
	\left[ e^{\frac{i(\omega -\vec{k}_{\mu s}\vec{v})L}
		{c\gamma _{0}}}-1\right] \right| ^{2},
\label{eq:99_11}
\end{equation}
here $\vec{e}_{1}||\left[ \vec{k}\vec{\tau}\right] $ , $\vec{e}_{2}||\left[
\vec{k}\vec{e}_{1}\right] $ , $\vec{k}_{\mu s}=\vec{k}+\frac{\omega }{%
	c\gamma _{0}}\varepsilon _{\mu s}\vec{N}$.

For PXR quanta with polarization vector $\vec{e}_{\tau s}$  diffracted at
a large angle relatively to the  particle velocity the above
formula should be rewritten as follows:
\begin{equation}
	\frac{d^{2}N_{s}}{d\omega d\Omega }=\frac{e^{2}Q^{2}\omega }{4\pi
		^{2}\hbar c^{3}}(\vec{e}_{\tau s}\vec{v})^{2}
	\left| \sum_{\mu =1,2}\gamma _{\mu s}^{\tau }\left[ \frac{1}{\omega -\vec{k}_{\tau }
		\vec{v}}-\frac{1}{\omega -\vec{k}_{\mu \tau s}\vec{v}}\right] \left[
	e^{\frac{i(\omega -\vec{k}_{\mu \tau s}\vec{v})L}{c\gamma _{0}}}-1\right]
	\right| ^{2}.
	\label{eq:80}
\end{equation}

\subsection*{A.2. Transition radiation and PXR}

Let us compare the obtained formulas with the well-known those for
transition X-ray radiation (TXR) in amorphous medium. For TXR the following expression is known:
\begin{equation}
	\frac{d^{2}N_{s}}{d\omega d\Omega } =\frac{e^{2}Q^{2}\omega }{4\pi
		^{2}\hbar c^{3}}(\vec{e}_{s}\vec{v})^{2}
	\left| e^{i\frac{\chi _{0}\omega }{2c\gamma _{0}}L}\left[ \frac{1}{
		\omega -\vec{k}\vec{v}}-\frac{1}{\omega -\vec{k}_{a}\vec{v}}\right] \left[
	e^{\frac{i(\omega -\vec{k}_{a}\vec{v})L}{c\gamma _{0}}}-1\right] \right|
	^{2},
\label{eq:99_13}
\end{equation}
where $\vec{k}_{a}=\vec{k}+\frac{\omega \chi _{0}}{2c\gamma _{0}}\vec{N}$ \
is the photon wave vector in amorphous medium.
The only single wave propagates in either amorphous medium or in a crystal away from the diffraction
conditions.
When diffraction takes place, then a coherent
superposition of several waves propagates in a crystal.
One can see that (\ref{eq:99_13}) is very much similar to the expression describing PXR.
When the frequencies and angles in (\ref{eq:99_11}) do not satisfy the
diffraction conditions, the formula for forward PXR turns to formula (\ref{eq:99_13}) for TXR. At the same time the spectral-angular intensity
(\ref{eq:80}) appears to be equal to zero, that is quite obvious
since no diffraction arises at a large angle.


Every item in (\ref{eq:80}) describes photon radiation amplitude $A_{\mu
	s}$ arising as a result of charged particle movement
through the crystal target of  thickness $L$. 
Since there are two
refraction indices, the total radiation density is expressed via
the squared module of amplitudes sum $\frac{d^2
	N_s}{d\omega d\vec{O}}\sim |A_{1s}+A_{2s}|^2$.

As $\chi'_0<0$, therefore, from the Vavilov-Cherenkov condition it
follows that the real part of
refraction index $n'>1$  for a single root only $(\mu=1)$.
As a result the difference $(\omega-\vec
k_{1\tau s}\vec v)$ can turn to zero and the term comprising this difference in a denominator in  (\ref{eq:80}) grows proportionally to $L$. 
At the first sight it means that the term,
comprising this difference (quasi-Cherenkov term)
mainly contributes into radiation, when the crystal thickness along the particle velocity is increased.
However, Bragg  diffraction case differs considerably from the case of Laue
diffraction, namely, in some range of frequencies and angles the
phenomenon of total reflection takes place. 
In this range of angles
and frequencies the wave vectors in crystal  are 
imaginary and wave is not  absorbed. 
In case of total reflection stipulated by the existence of the heterogeneous wave in the crystal, it is necessary to take into
account both dispersion branches, when calculating the
radiation intensity for  Bragg  diffraction scheme.
 Though the
structure of expression (\ref{eq:80}) is very simple, but in order to obtain
quantitative data it is necessary to performed correct
calculations following (\ref{eq:80}) taking into account all terms: presence of oscillating terms and their interference could 
cause wrong results if some neglected.

\end{document}